\documentclass[aps,prl,twocolumn,amsmat,amssymb,amsfonts,superscriptaddress]{revtex4-2}

\usepackage{amsmath, amsfonts}

\usepackage{hyperref}
\hypersetup{colorlinks=true,
		    linkcolor=blue,
		    citecolor=blue,
		    allcolors=blue}

\usepackage{color}
\usepackage{graphics, graphicx}

\usepackage{ifthen, times}
\usepackage{tikz}
\usetikzlibrary{arrows}
\usetikzlibrary{decorations.pathreplacing,decorations.markings}

\usepackage[percent]{overpic}
\usepackage{wasysym}

\makeatletter
\g@addto@macro\bfseries{\boldmath}
\makeatother

\usepackage{bm, bbm, braket, mathtools, comment}
\usepackage{tcolorbox}
\tcbuselibrary{theorems}
\renewcommand{\H}{\mathcal{H}}

\newcommand{\Minus}{\mathord{\tikz\draw[line width=0.3ex, x=1ex, y=1ex] (0.0,1.0) -- (1.,1.0);}}
\newcommand{\xbond}{{\color{blue}\Minus}}
\newcommand{\ybond}{{\color{red}\Minus}}

\newcommand{\shalf}{$s$=$\tfrac{1}{2}$}

\begin{document}

\title{Field Induced Chiral Soliton Phase in the Kitaev Spin Chain}
\author{Erik S. S{\o}rensen}
\email{sorensen@mcmaster.ca}
\affiliation{Department of Physics \& Astronomy, McMaster University, Hamilton ON L8S 4M1, Canada.}
\author{Jacob Gordon}
\affiliation{Department of Physics, University of Toronto, Ontario M5S 1A7, Canada}
\author{Jonathon Riddell}
\affiliation{Department of Physics \& Astronomy, McMaster University, Hamilton ON L8S 4M1, Canada.}
\author{Tianyi Wang}
\affiliation{Department of Physics, University of Toronto, Ontario M5S 1A7, Canada}
\author{Hae-Young Kee}
\email{hykee@physics.utoronto.ca}
\affiliation{Department of Physics, University of Toronto, Ontario M5S 1A7, Canada}
\affiliation{Canadian Institute for Advanced Research, CIFAR Program in Quantum Materials, Toronto, ON M5G 1M1, Canada}
\date{\today}

\begin{abstract}
The bond-dependent Ising interaction present in the Kitaev model has attracted
considerable attention. The appearance of an unexpected intermediate phase
under a magnetic field is particularly intriguing, and one may wonder if a
similar phase occurs in the Kitaev spin chain with alternating $x$- and
$y$-bond Ising interactions. Previous studies have focused on a transverse
field, $h_z$, and reported a direct transition to the polarized state. Here,
we investigate phases with arbitrary angle of two longitudinal fields, $h_x$
and $h_y$.  For a magnetic field applied along the diagonal, $h_x$=$h_y$, the
chain remains gapless up to a critical field $h^{c_1}_{xy}$.  Surprisingly,
above $h^{c1}_{xy}$ it enters an unusual intermediate phase before reaching
the polarized state at $h^{c_2}_{xy}$.  This phase is characterized by a
staggered vector chirality and for periodic boundary conditions, a two-fold
degeneracy with a finite gap.  For open boundary systems the ground-state
exhibits a single {\it soliton}, lowering the energy, and gapless excitations.  However, the
corresponding anti-soliton raises the energy sufficiently that a gap appears
for soliton and anti-soliton  pairs in periodic systems.  An intuitive
variational picture is developed describing the soliton phase.
\end{abstract} 
\maketitle

{\it Introduction}: The Kitaev model, characterized by the bond-dependent Ising spin interaction in the honeycomb lattice~\cite{kitaev2006}, has recently generated considerable interest, as it offers a rare quantum spin liquid as an exact ground-state. 
Among several exotic phenomena discussed in relation to the extended Kitaev model~\cite{rau2014prl}, the proposed field-induced $U(1)$ spin liquid in the antiferromagnetic (AFM) Kitaev model under a magnetic field is especially fascinating~\cite{hickey2019visons,lu2018spinon,zou2020neutral}. 
While the mechanism of the $U(1)$ spin liquid is still missing, a magnetically disordered phase featuring a staggered scalar chirality has been found in the quasi-one-dimensional AFM Kitaev ladder under a [111] magnetic field~\cite{Sorensen2021}.
The relation between these phases, if any, is at present not clear and a detailed understanding of the field dependent phase diagram as the two-dimensional limit is approached, starting from the purely one-dimensional (1D) Kitaev spin chain, would clearly be desirable. This then raises the question if any non-trivial phases exists for the Kitaev spin chain in a magnetic field.

The 1D Kitaev \shalf\ spin chain, 
has been investigated and shown to map to a free fermion model~\cite{Feng2007topological,Sun2009,Wang2010,You2018,Wu2019}.
With the Kitaev chain defined in terms of $x-$ and $y-$bond Ising interactions, it has been shown that under
a transverse magnetic field $h_z$, the model directly enters the polarized state without any phase transition~\cite{Sun2009}. In fact, so far no intermediate phase in an applied field has been reported for the Kitaev spin chain. 
One may then wonder if the Kitaev spin chain exhibits any intermediate phase under a magnetic field like the ladder and C$_3$ symmetric honeycomb lattice mentioned above. Here we address this question, and report an unusual intermediate chiral phase possessing magnetic solitons in the AFM Kitaev spin chain under a magnetic field close to the $h_x \sim h_y$. This phase is absent in the ferromagnetic (FM) Kitaev spin chain.
\begin{figure}
    \includegraphics[width=\columnwidth]{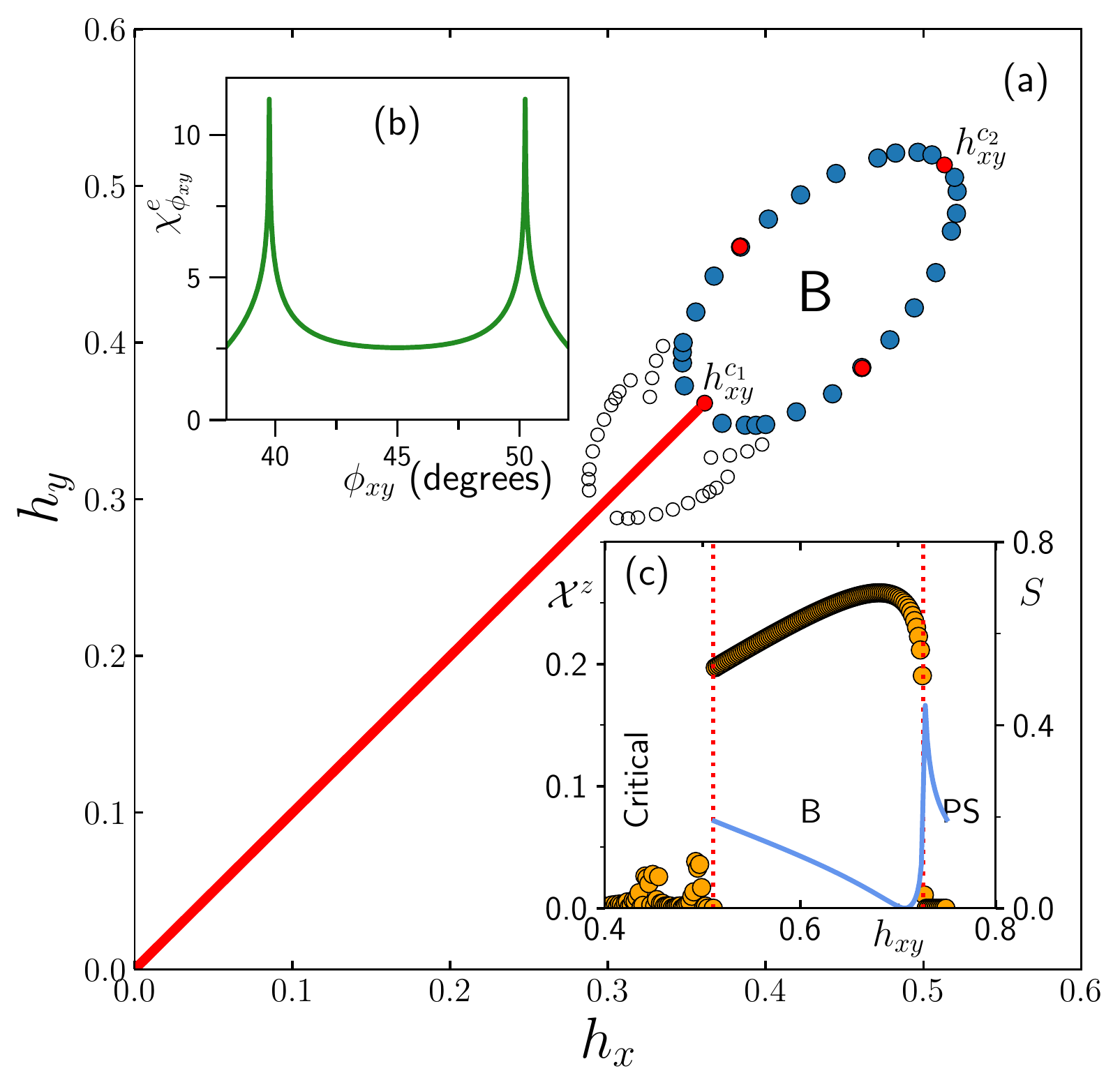}
    \caption{
    (a) The phase diagram in the $(h_x,h_y)$ plane from ED, $N=24$ (solid blue points), iDMRG (solid red points). The soliton phase is marked by `B'. Open circles indicate cross-over observed in ED due to incommensurability effects. 
    (b) iDMRG results for $\chi^e_{\phi_{xy}}$ at $|h|=0.6$ versus $\phi_{xy}$.
    (c) $\mathcal{X}^z$ versus $h_{xy}$ at a field angle to 45$^\circ$ from iDMRG (orange points) and the bipartite entanglement entropy $S$ (blue line).
    }  
    \label{fig:bigfig}
\end{figure}

The \shalf\ Kitaev spin chain is described by the Hamiltonian:
\begin{equation}\label{eq:H}
\H = K\sum_{j}\left(S_{2j+1}^xS_{2j+2}^x + S_{2j+2}^yS_{2j+3}^y\right)- \sum \bm{h} \cdot \bm{S}
\end{equation}
Where we set $g$=$\hbar$=$\mu_B$=$1$ and consider the AF model with $K$=1 and a  parameterization of the field term as
$\bm{h}$=$h(\cos\phi_{xy}\cos\theta_z,\sin\phi_{xy}\cos\theta_z,\sin\theta_z)$. We refer to the coupling $KS^xS^x$ as an $x$-bond ($\xbond$) and $KS^yS^y$ as a $y$-bond ($\ybond$).
Below, we determine the phase diagram of Eq.~(\ref{eq:H}) in a field using Lanczos exact diagonalization (ED) techniques in combination with DMRG and iDMRG~\cite{White1992b,McCulloch2008,Schollwock2011,itensor} methods typically performed with a bond dimension larger than 1,000 and a $\epsilon<10^{-10}$. Subsequently, we describe our variational calculations valid in the chiral soliton phase. 
\begin{figure}[t]
  \includegraphics[width=\columnwidth]{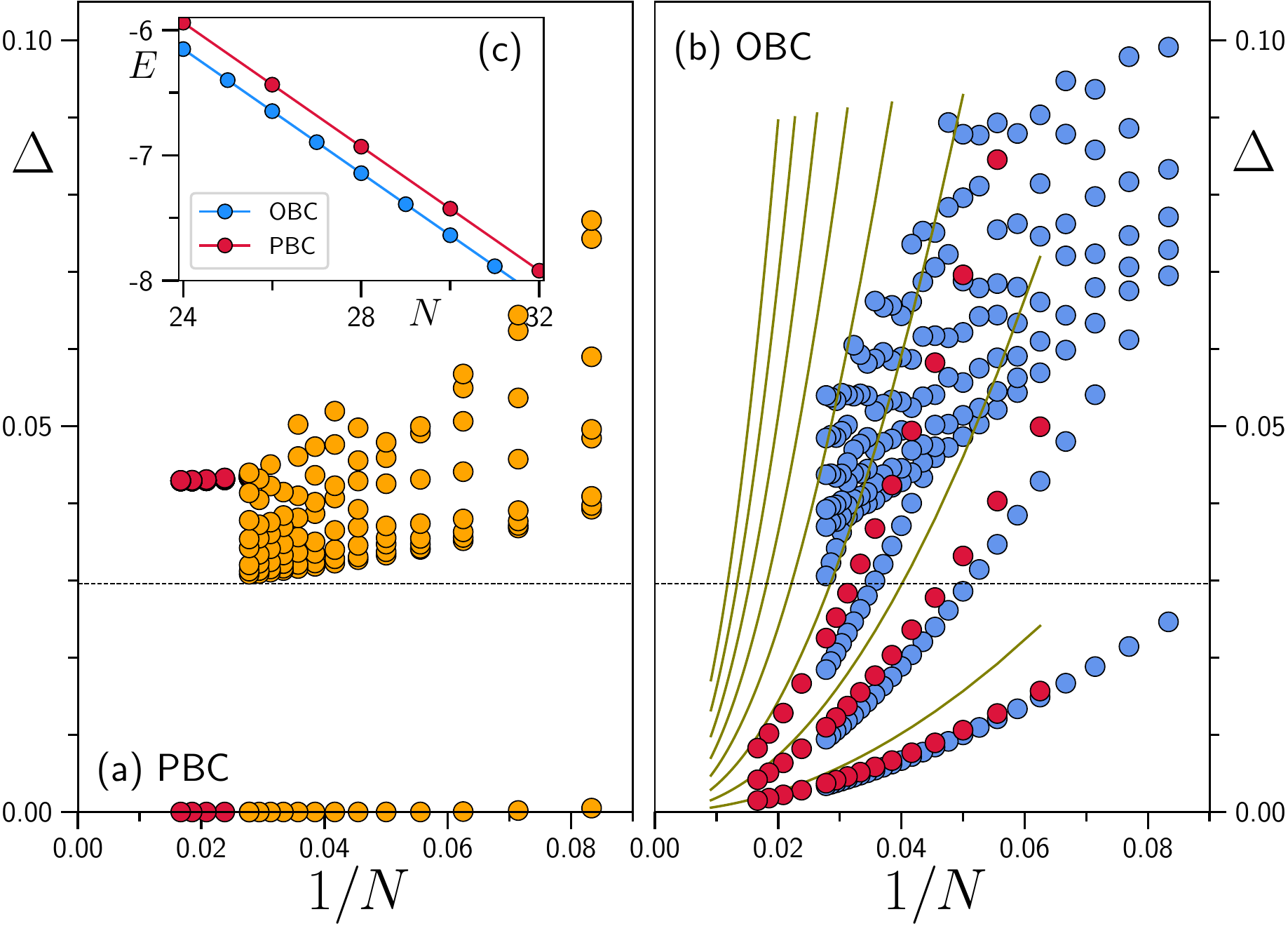}
  \caption{
     Energy gap $\Delta$ to the first 10 energy levels  above the ground-state (not shown) versus $1/N$ at $h_{xy}=0.7$. 
    (a) ED results with PBC for even $N=12$-$36$ (orange points). Note the twice degenerate ground-state below a well defined gap. The dashed line indicates $\Delta_{\rm PBC}=0.02962K$. Two soliton variational results (red points).
    (b) ED results with OBC for $N=12-36$ (blue points). Variational estimates for the lowest gaps in the space of single defects (green lines) and with $\{0,1,2\}$ defects (red points).
    (c) ED results for the ground-state energy versus $N$ at $h_{xy}$=0.7, for OBC (blue) and PBC (red). 
  }
  \label{fig:Gaps}
\end{figure}  

{\it Phase Diagram}: In the presence of a  field in the $z$-direction the Kitaev chain, Eq.~(\ref{eq:H}) is exactly solvable and it is known that the system immediately enters the polarized state (PS)~\cite{Sun2009} directly. The integrability is lost when the field is applied in the $x$- or equivalently the $y$-direction and the situation is less clear.  
We have therefore studied the correlation functions $C(r)$=$\langle S^x_1S^x_{r+1}\rangle$ for small fields in the $x$-direction.
For $h_x$=$0$ a power-law is found, $C(r)$$\sim$$r^{-0.25(1)}$, as shown in~\cite{SM}, however, for any non-zero $h_x$ an exponential decay is observed with a resulting finite gap~\cite{SM}.
The polarized state is then entered directly for any non-zero $h_x$ and by symmetry for any non-zero $h_y$.

Next we study the phase diagram for fields in the entire $x$-$y$ plane, Fig.~\ref{fig:bigfig}(a). Although difficult to establish numerically, our results indicate that for $\phi_{xy}$=$45^\circ$ the Kitaev chain remain gapless up to a critical field, $h^{c_1}_{xy}$=$0.511K$ where a new unexpected phase is entered, marked by `B'. 
We determine the phase boundary for this phase by studying the energy susceptibility 
$\chi^e_{\phi_{xy}}$=$-\partial^2 e_0/\partial\phi_{xy}^2$ 
which scales as $N^{2/\nu-(d+z)}$ at a quantum critical point~\cite{Albuquerque2010}. 
Here, $e_0$ is the energy per site and $\nu,z$ the correlation and dynamical exponents. The solid blue points denote results from $N$=$24$ ED where $\chi^e_{\phi_{xy}}$ is maximal. 
The position of these peaks are confirmed by iDMRG (solid red points) as illustrated in the inset, Fig.~\ref{fig:bigfig}(b). The open circles denote crossover due to incommensurability effects where the position of the ED peak cannot be reproduced with iDMRG and is strongly finite-size dependent. The phase extends out of the $x$-$y$ plane to non-zero $\theta_z$\cite{SM}. 
At a 45$^\circ$ angle another quantum critical point is observed at the critical field $h^{c_2}_{xy}$=$0.726K$ where the chain transitions from the soliton phase to the polarized state. 
\begin{figure}
  \includegraphics[width=\columnwidth]{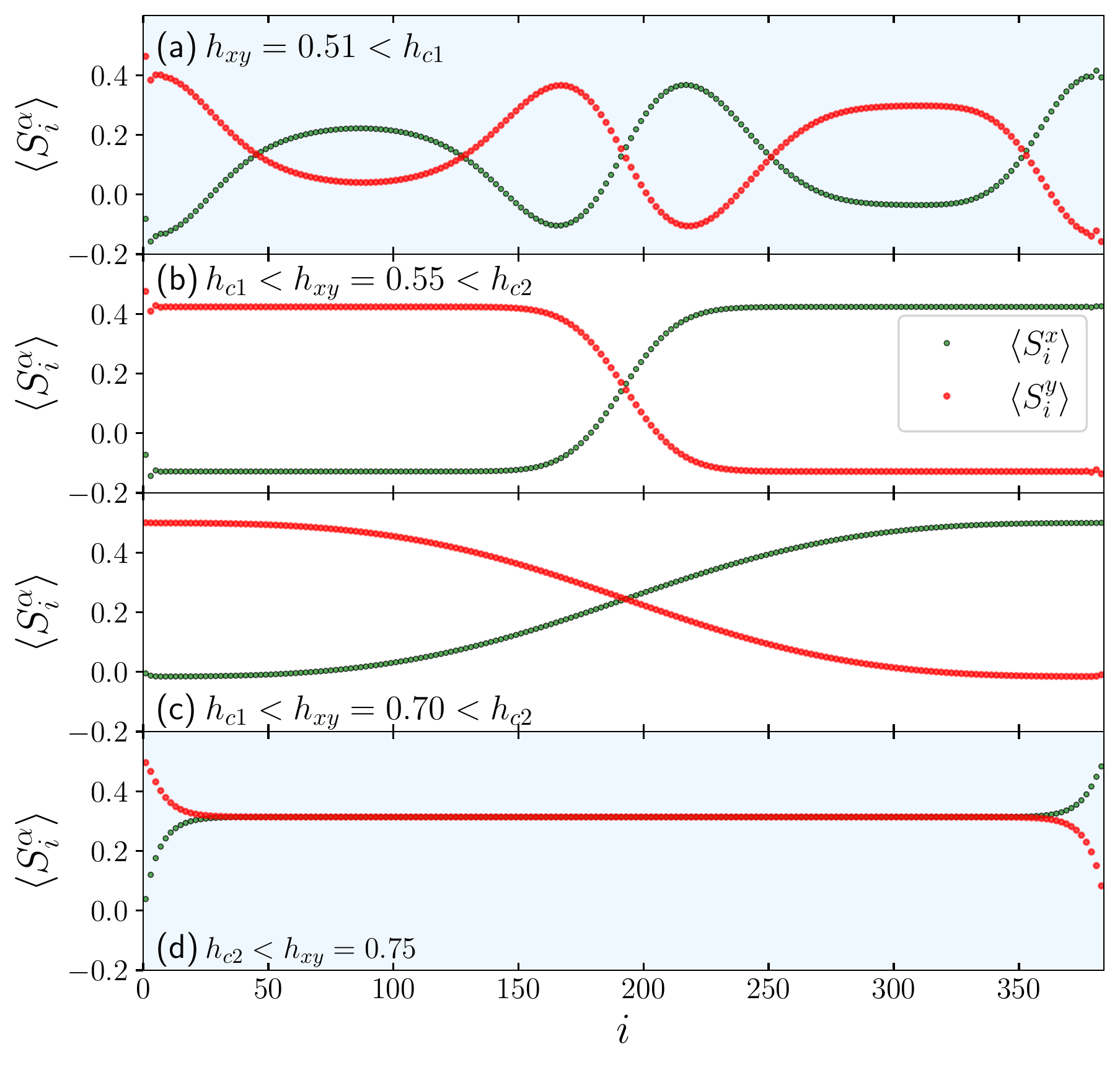}
  \caption{
    (a)-(d) Finite chain DMRG results for the on-site magnetization $\langle S^x_i \rangle$, $\langle S^y_i \rangle$ versus position, $i$, for a $N=384$ site chain with OBC at different field strengths. (a) $h_{xy}=0.51<h_{c1}$, (b) $h_{xy}=0.55$
    (c) $h_{xy}=0.70$ (d) $h_{xy}=0.75>h^{c_2}_{xy}$. Only {\it odd} sites are shown.
  }
  \label{fig:Butterflies}
\end{figure}  

The gapless phase, extending from zero field to $h^{c_1}_{xy}$ at a 45$^\circ$ angle, is a critical 
{\it line}. For $\phi_{xy}\neq 45^\circ$, or $\theta_z\neq 0$, a gap opens up and the chain enters the PS phase. 
The B phase is characterized by a non-zero staggered vector chirality, $\mathcal{X}^\alpha$: 
\begin{equation}
    \mathcal{X}^\alpha= (-1)^j\langle ({\bf S}_j \times {\bf S}_{j+1})^\alpha\rangle.
\end{equation}
While ${\mathcal{X}}^{x,y}$=$0$ in the B phase ${\mathcal{X}}^z$$\neq$0 as shown in Fig.~\ref{fig:bigfig}(c).
In the context of of the anisotropic $J_1$-$J_2$ model with $J_1<0$, $J_2>0$~\cite{Furukawa2008,Furukawa2010,Furukawa2012} phases with non-zero $\mathcal{X}^\alpha$ have been found and recently observed in the \shalf\ chain LiCuVO$_4$~\cite{Grams2022}.
\begin{figure}
  \includegraphics[width=\columnwidth]{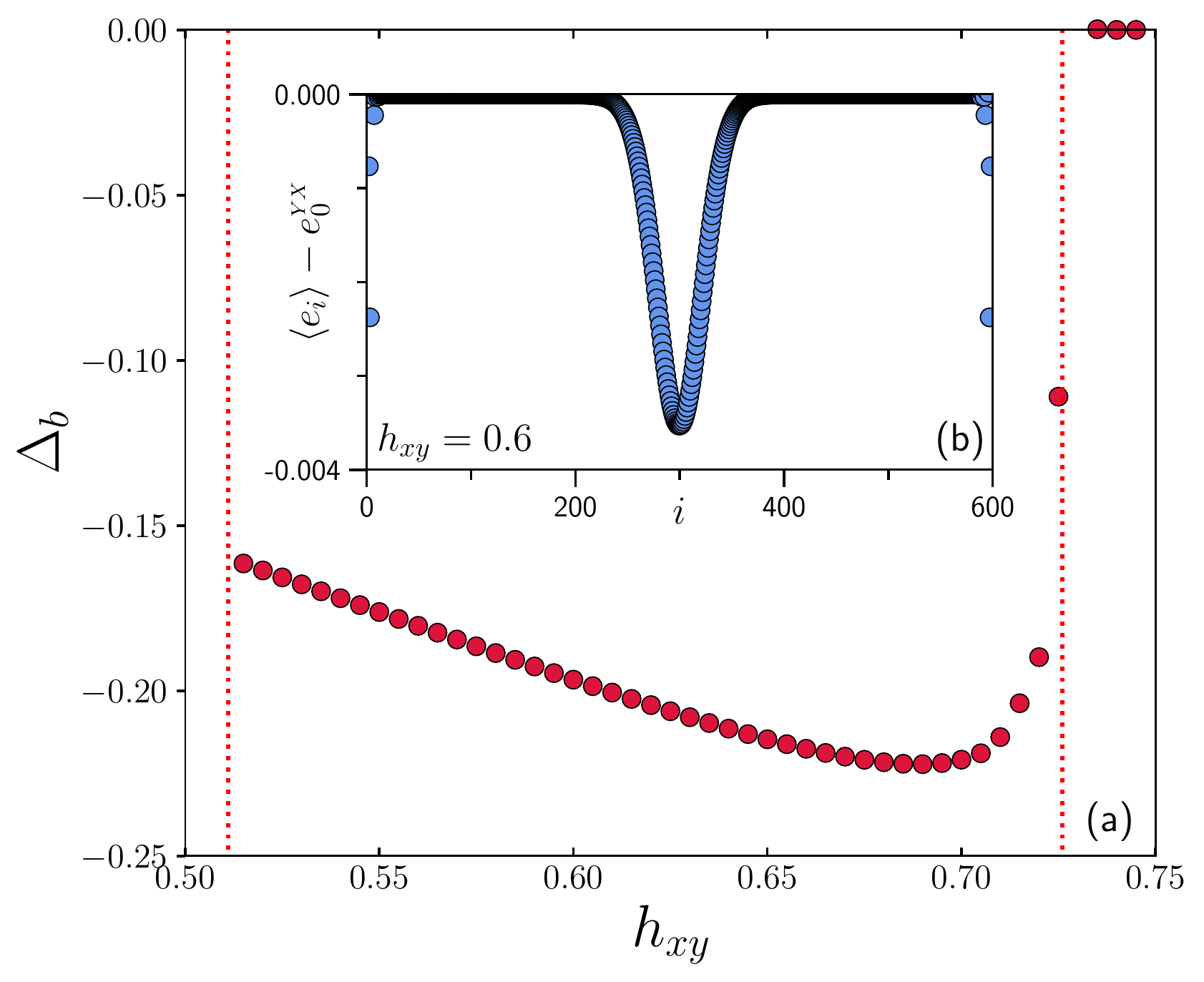}
  \caption{
    (a) $N$=$1200$ DMRG results with OBC for the
    soliton mass, $\Delta_b$ versus $h_{xy}$ at $\phi_{xy}=45^\circ$. 
    (b) $\langle e_i\rangle$-$e_0^{\scriptscriptstyle YX}$
    versus $i$ at $h_{xy}=0.6$, $\phi_{xy}=45^\circ$.
  }
  \label{fig:Deltab}
\end{figure}  

To understand the nature of the unexpected B phase we focus on the spectrum of excitations. Using ED for chain sizes ranging from $N$=$12$-$36$ at $h_{xy}$=$0.7, \phi_{xy}$=$45^\circ, \theta_z$=$0^\circ$ results for the
gap to the 10 lowest states  are shown in Fig.~\ref{fig:Gaps}(a) for PBC (orange points) and Fig.~\ref{fig:Gaps}(b) for OBC (blue points). For PBC there are {\it two} almost degenerate states that become degenerate as $N\to\infty$ below a well defined but small gap of $\Delta_{\rm PBC}$=$0.02962K$. For PBC we determine the momentum of the lowest excited state above the 2 degenerate ground-states to be at $k=0$. 
For OBC the spectrum is more intriguing. As seen in Fig.~\ref{fig:Gaps}(b) the spectrum evolves smoothly with $N$ for both even and odd $N$. While it is possible to identify $\Delta_{\rm PBC}$ in the spectrum for OBC an increasing number of states appear {\it below} this gap, quickly approaching the ground-state energy. Counter-intuitively, as shown in Fig.~\ref{fig:Gaps}(c) the ground-state energy is {\it always lower} for OBC as compared to PBC for any $N$, despite the missing bond.
At $h_{xy}$=$0.7$ we determine $\Delta_\mathrm{O-PBC}$=$-0.2121K$. Open boundary conditions therefore allow the chain to significantly lower the energy. The proliferation of states below the gap for OBC is an unusual feature reflecting excited states of the soliton, as we discuss below.

The ground-state magnetization $\langle S^\alpha_i\rangle$ is very unique in the B phase for OBC. As seen in Fig.~\ref{fig:Butterflies}(b)-(c), $\langle S^\alpha_i\rangle$ (shown only for {\it odd} sites) alternates between $x$- and $y$-directions with a single twist, a {\it topological soliton}, occuring in the middle of the chain. 
For $h_{xy}<h^{c_1}_{xy}$ the onsite magnetization is much more complicated and 5 crossings are present for $N=384$. In the PS phase for $h_{xy}>h^{c_2}_{xy}$ the spins simply align with the field and the soliton is absent. A useful way to visualize
the solitons is to plot the energy density for each bond $e_i$. In the bulk this is just a constant,
$e_0^{\scriptscriptstyle YX}$, but the presence of the soliton {\it lowers} $e_i$ below this value, locally.
This is shown in Fig.~\ref{fig:Deltab}(b) where $\langle e_i\rangle-e_0^{\scriptscriptstyle YX}$ is plotted versus $i$ for $h_{xy}=0.6$, showing a sharply localized soliton. If we now evaluate:
\begin{equation}
\Delta_b=\sum_i \left( \langle e_i\rangle-e_0^{\scriptscriptstyle YX}\right),\label{eq:sumei}
\end{equation}
we can determine by how much the soliton has lowered the total energy which we denote the soliton mass, $\Delta_b$. Results for
$\Delta_b$ calculated this way are shown in Fig.~\ref{fig:Deltab}(a) throughout the soliton phase. While closely related,
$\Delta_\mathrm{O-PBC}$ includes boundary effects from the missing bond with OBC.

{\it Variational Picture, PBC:}
As shown in Fig.~\ref{fig:bigfig}(c) the entanglement entropy $S$ is relatively low in the soliton phase. In fact,
for PBC and $N$ even the two-fold degeneracy noted in the ground-state subspace in Fig.~\ref{fig:Gaps}(a) is closely described by two (zero-defect) {\it product} states of the following form:
\begin{equation}
    |X'Y'\rangle=|x'y'x'y'\ldots\rangle,\ \ |Y'X'\rangle=|y'x'y'x'\ldots\rangle,
\end{equation}
where $|y'\rangle =(e^{-i(\pi/2+c)},1)/{\sqrt{2}},$
$|x'\rangle =(e^{ic},1)/{\sqrt{2}}.$ The $|x'\rangle$ and $|y'\rangle$ are 
eigen-states of $\vec S\cdot \vec n_{\alpha}$ where the unit vectors $\vec n_{x'}, \vec n_{y'}$ 
are close to the $x-$ and $y-$ direction but {\it crucially} with an angle between them exceeding $\pi/2$, by $2c$.
The usual $|x\rangle$ and $|y\rangle$ states are obtained by setting $c=0$.
The optimal value for $c$ depends on the field $h_{xy}$ and is determined in~\cite{SM} to be:
$c=\cos^{-1}(h_{xy}/K)$-$\pi/4$.
The solitons shown in Fig.~\ref{fig:Butterflies}(b)-(c) for OBC then interpolate between these two degenerate states as is typical for {\it topological} solitons~\cite{Solitons}.
Although $\langle X'Y'|H|Y'X'\rangle$ is non-zero for very short chains this coupling quickly goes to zero with $N$.

{\it OBC, any N:} We now focus on OBC irrespective of $N$,
and we focus exclusively on the case where the chain {\it starts with} a $S^x_1S^x_2$ term ($\xbond$), in which case the solitons in Fig.~\ref{fig:Butterflies} transition from the $y'x'$ to the $x'y'$ pattern. Within the soliton phase the lowest energy subspace is well described 
by linear combinations of (single defect) states of the form:
\begin{eqnarray}
    |\psi_b(i)\rangle=|
    y'\xbond\ x'\ybond\ 
    y'\xbond\ 
    \tcbset{colback=blue!10!white}
    \tcboxmath[size=fbox,auto outer arc, arc=5pt]{
    x'_i\ybond\ x'
    }
    \xbond\ y'\ybond\
    x'\xbond\ y'\ybond\
    x'\xbond\ y'\rangle,\nonumber\\
    |\psi_b(i)\rangle=|
    y'\xbond\ x'\ybond\ 
    y'\xbond\ x'\ybond\
    \tcbset{colback=blue!10!white}
    \tcboxmath[size=fbox,auto outer arc, arc=5pt]{
    y'_i\xbond\ y'
    }
    \ybond\
    x'\xbond\ y'\ybond\
    x'\xbond\ y'\rangle,
    \label{eq:psib}
\end{eqnarray}
transitioning from the $y'x'$ to the $x'y'$ pattern at bond $i$. Note that, even though 
$ x'_i\ybond\ x' $ have the spins aligned ferromagnetically along $x$, it costs little energy since it occurs on a $y$-bond. Similarly,
the spins at
$ y'_i\xbond\ y' $ are aligned ferromagnetically along $y$, but on a $x$-bond.
Analoguously, we can define `anti'-defects of the form
\begin{eqnarray}
    |\psi_B(i)\rangle=|
    x'\xbond\ y'\ybond\ 
    x'\xbond\ 
    \tcbset{colback=red!10!white}
    \tcboxmath[size=fbox,auto outer arc, arc=5pt]{
    y'_i\ybond\ y'
    }
    \xbond\ x'\ybond\
    y'\xbond\ x'\ybond\
    y'\xbond\ x'\rangle,\nonumber\\
    |\psi_B(i)\rangle=|
    x'\xbond\ y'\ybond\ 
    x'\xbond\ y'\ybond\
    \tcbset{colback=red!10!white}
    \tcboxmath[size=fbox,auto outer arc, arc=5pt]{
    x'_i\xbond\ x'
    }
    \ybond\
    y'\xbond\ x'\ybond\
    y'\xbond\ x'\rangle,\label{eq:psiB}
\end{eqnarray}
in this case transitioning from the $x'y'$ to the $y'x'$ pattern at bond $i$. Contrary to the defects, $\psi_b$, these anti-defects are relatively {\it costly} since $y'_i\ybond\ y'$ now occurs on a $y$-bond and $x'_i\xbond\ x'$ on a $x$-bond.

The states $\psi_b$ and $\psi_B$ are not eigenstates of the Hamiltonian~\cite{SM} but we expect linear combinations of the single defect $|\psi_b(i)\rangle$ 
to realistically define the low-energy sub-space of the system in a variational manner. We therefore define the (single defect) soliton states:
\begin{equation}
    |\Psi_b\rangle=\sum_{k}a_k|\psi_b(k)\rangle,
    \label{eq:Psib}
\end{equation}
Similarly, we can define $|\Psi_B\rangle=\sum{c}_l|\psi_B(l)\rangle$ but this leads to {\it high} energy states.
It is important to note that the states $|\psi_b(i)\rangle$, while normalized, are not orthogonal. 
Due to the non-orthogonality, determining the optimal values for the coefficients $a_k$ in $\Psi_b$ in Eq.~(\ref{eq:Psib}) 
from a variational calculation therefore defines a generalized eigenvalue problem in terms of the matrices ${\cal H}_{kl}=\langle \psi_b(k)|H|\psi_b(l)\rangle$ and ${\cal S}_{kl}=\langle \psi_b(k)|\psi_b(l)\rangle$, with similar definitions for the state $\Psi_B$.  The generalized eigenvalue problem can be solved using standard methods and the optimal $\Psi_b$ and $\Psi_B$ determined.
\begin{figure}
  \includegraphics[width=\columnwidth]{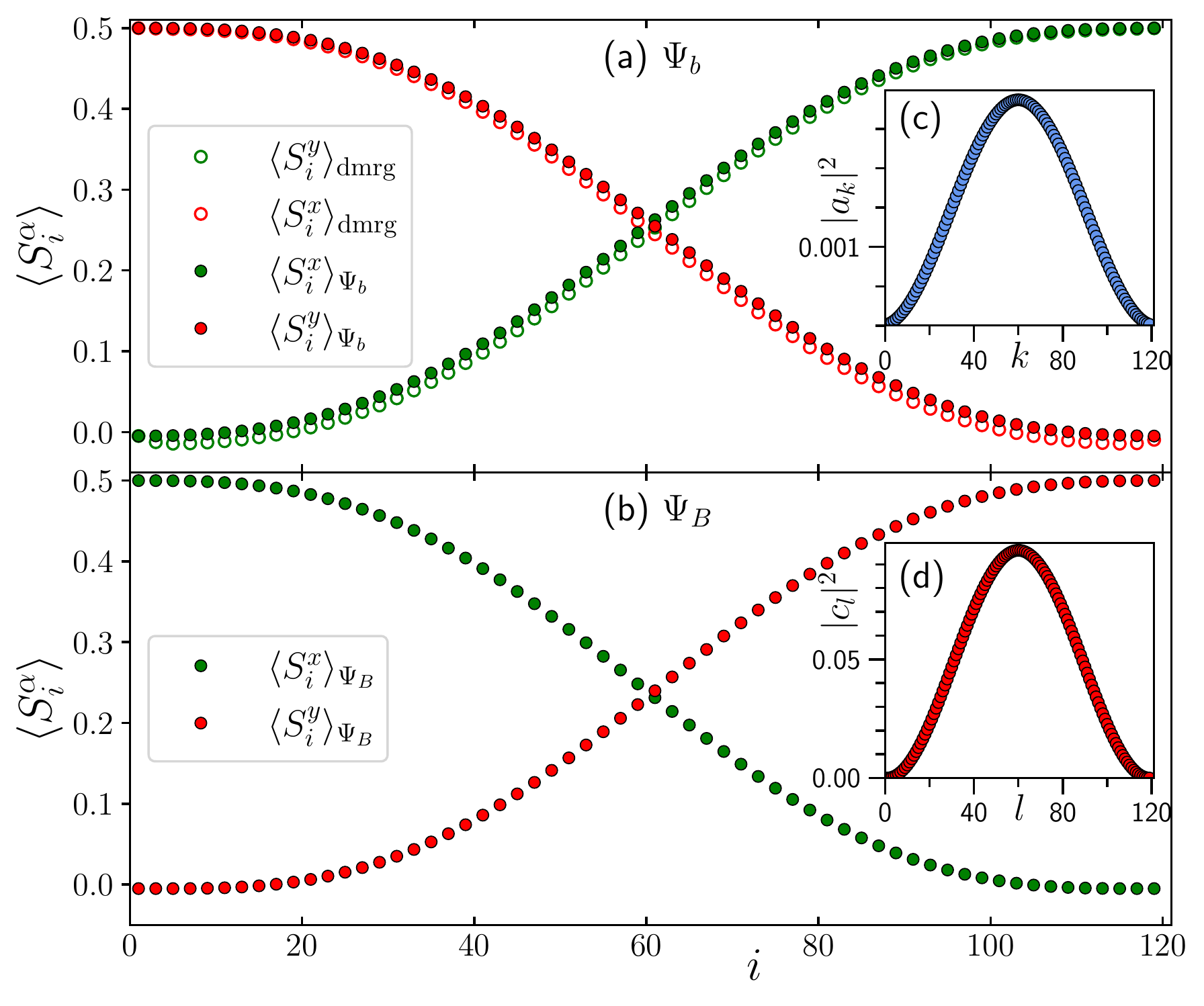}
  \caption{
    (a)$\langle S^x_i\rangle$ and $\langle S^y_i\rangle$ on {\it odd} sites from DMRG calculations on a $N=120$ site open chain compared to the 
    variational results $\langle S^x_i\rangle_{\Psi_b}$, $\langle S^x_i\rangle_{\Psi_b}$ in the variational soliton state $\Psi_b$ at $h_{xy}=0.7$, $\phi_{xy}$=$45^\circ$.
    (b)$\langle S^x_i\rangle_{\Psi_B}$, $\langle S^x_i\rangle_{\Psi_B}$ on {\it odd} sites in the variational `anti'-soliton state $\Psi_B$. 
    (c) $|a_k|^2$ versus $k$ for the $\Psi_b$ state.
    (d) $|c_l|^2$ versus $l$ for the $\Psi_B$ state.
  }
  \label{fig:Variational}
\end{figure}  

{\it Variational Results}:
Solving the generalized eigenvalue problem yields a series of states for $\Psi_b$ and $\Psi_B$. With OBC we expect the lowest $\Psi_b$ state to be a good approximation to the ground-state. This is illustrated in Fig.~\ref{fig:Variational}(a),(c)
where the variational results for $\langle S^{x,y}_i\rangle_{\Psi_b}$ are compared to finite chain DMRG results for a system with $N$=$120$ at $h_{xy}$=$0.7$. 
We find $E_{\rm DMRG}$=$-29.9169$ while $E_{\Psi_b}$=$-29.9019$ less than 0.05\% higher. For comparison, the defect free $Y'X'$ state has an energy $E_{Y'X'}$=$-29.6975$ significantly higher and the soliton has therefore {\it lowered} the energy with respect to the $Y'X'$ state. However, for the `anti'-soliton state $\Psi_B$ shown in Fig.~\ref{fig:Variational}(b),(d) we instead find
$E_{\Psi_B}$=$-29.4520$ {\it above} the $Y'X'$ state. 
Using the defect free $Y'X'$ state as reference we can now estimate the energy difference (mass) for the two states at $h_{xy}$=$0.7$:
$    \Delta_b$=$-0.2044 K$ compared to $-0.2085K$ from DMRG in Fig.~\ref{fig:Deltab}(a) and $\Delta_B$=$0.2455K$
which cannot be determined from DMRG nor ED.
A similar asymmetry has been noted in the Rice-Mele model~\cite{Allen2022} and the nonsymmomorphic symmetry~\cite{Wang2022} present also in the Kitaev spin chain could be crucial. 

For PBC the ground-state in the soliton phase is well described by the degenerate and defect-free $Y'X'$ and $X'Y'$ states. While for OBC the number of solitons, $n_{\rm sol}$, can be both even and odd, with PBC it is not possible to consider a single soliton, they have to come in a $bB$ pair or multiple pairs, $0$,$bB$,$bBbB$,$\ldots$, with $n_{\rm sol}$ {\it even}. 
This explains the gap seen in Fig~\ref{fig:Gaps}(a) since to a first approximation we expect that
\begin{equation}
    \Delta_{\rm PBC}=\Delta^b+\Delta^B,
\end{equation}
which would predict a gap for PBC of $0.0411K$ from the variational results. For OBC the gap to the lowest $bBb$ state from the $b$ ground-state should then also be equal to $\Delta_{\rm PBC}$ which
agrees with the results in Fig~\ref{fig:Gaps}(b).
We then extend the variational calculations to two-defect $bB$ states by considering:
\begin{equation}
    |\psi_{bB}(i,j)\rangle=|
    y'\xbond\ 
    \tcbset{colback=blue!10!white}
    \tcboxmath[size=fbox,auto outer arc, arc=5pt]{
    x'_i\ybond\ x'
    }
    \xbond\ y'\ybond\ x'\xbond y'\ybond
    \tcbset{colback=red!10!white}
    \tcboxmath[size=fbox,auto outer arc, arc=5pt]{
    x'_j\xbond\ x'
    }
    \ybond y'\xbond x'
    \rangle,\label{eq:psibB}
\end{equation}
and defining two-soliton states of the form:
\begin{equation}
|\Psi_{bB}\rangle=\sum_{i\neq j} a_{i,j}|\psi_{bB}(i,j)\rangle.
\end{equation}
If we include the $Y'X'$ and $X'Y'$ states in the variational calculation, extending the subspace to $\{0,2\}$ defects, we find  at $h_{xy}$=$0.7$ a gap of $\Delta_{PBC}^{var}$=$0.04289K$ (red circles in Fig.~\ref{fig:Gaps}(a)), in qualitative agreement with the ED result of $\Delta_{\rm PBC}$=$0.02962K$ and close to $\Delta_b$+$\Delta_B$.
We expect the inclusion of multiple pairs of defects in the variational subspace to further improve the agreement.
We can now intuitively understand the transition at $h^{c_1}_{xy}$. At this point $\Delta_b$=$-\Delta_B$ and the cost of a $bB$ pair becomes zero. As is clearly seen in Fig.~\ref{fig:Butterflies}(a) a number of $bB$ pairs then condense into the single soliton ground-state in this case creating a $bBbBb$ state. As the field is increased the solitons then effectively {\it evaporate}.
On the other hand, the transition at $h^{c_2}_{xy}$ occurs due to the closing of the gap to spin-wave excitations.

The solution of the generalized eigenvalue problem leads not only to the variational ground-state $\Psi_b$ but also a series of excitations of these states, $^i\Psi_b$ which are in good agreement with results for excited states obtained from DMRG~\cite{SM}.
For OBC, these states correspond to static excitations of the single soliton present in the system~\cite{Rajaraman}. As the system size
is increased the excited states  gradually fill in the gap in the spectrum. 
The variationally determined gaps obtained from the single defect states, Eq.~(\ref{eq:psib})  are shown as the green lines
in Fig.~\ref{fig:Gaps}(b). 
For short chains with OBC we can extend the variational subspace in Eq.~(\ref{eq:Psib}) to include
$\{0,1,2\}$ defects with the resulting gaps shown as red circles in Fig.~\ref{fig:Gaps}(b) significantly improving the agreement with the ED results for short chains.

{\it Discussion}:  In parallel 
with studies of solitons in conducting polymers~\cite{Heeger1988}, magnetic solitons have been studied since the late seventies~\cite{Mikeska1978,Mikeska1980,Fogedby1980a,Fogedby1980b,Kosevich1990,Mikeska1991} with signatures observed experimentally~\cite{Kjems1978} in the 1D easy-plane ferromagnetic chain system CsNiF$_3$ as well as the 1D AF materials TMMC~\cite{Boucher1985,Regnault1982}, CsCoBr$_3$~\cite{Buyers1986} and CsMnBr$_3$~\cite{Gaulin1985} among others.
The excitations of interest here are topological solitons linking distinguishable ground-states~\cite{Solitons}.
Building on this picture, domain walls between degenerate ground-states in dimerized spin chains, such as the \shalf, $J_1$-$J_2$ model, have been viewed as solitons~\cite{Shastry1981,Caspers82,Caspers84,Sorensen1998,Sorensen2007a,Sorensen2007b} and observed experimentally in BiCu$_2$PO$_6$ above a critical field~\cite{Casola2013}. Comparing periodic (PBC) and open (OBC) boundary conditions, a {\it positive} mass, $\Delta_s$, has then been defined~\cite{Sorensen1998,Sorensen2007a,Sorensen2007b} 
for both the soliton and anti-soliton in the dimerized phase with well defined spin, \shalf.

In contrast, for the Kitaev chain we find here that the soliton mass, $\Delta_b$ is {\it negative}, lowering the energy in the soliton phase, 
while the anti-soliton has {\it positive} mass,  raising
the energy by $\Delta_B$, more than compensating the soliton.
In periodic systems, the low lying excitation, a pair of soliton and anti-soliton then has a small gap given by the difference between $\Delta_B$ and $\Delta_b$. 
At $h^{c1}_{xy}$ the two masses cancels out, $\Delta_b$=$-\Delta_B$.
Furthermore, the soliton and anti-soliton do not have well defined spin.

Several important tasks remain to be addressed in future work.
One is finding candidate Kitaev spin chain systems.
Recently CoNb$_2$O$_6$ was proposed as a twisted Kitaev chain~\cite{Morris2021}.
However, the Kitaev interaction is FM and finding an AFM sister material would be of considerable interest. Preliminary results for the AFM twisted Kitaev chain show that the phase diagram is similar to the Kitaev chain considered here~\cite{SM}.
Furthermore, the presence of solitons should have important implications for thermodynamic properties such as the specific heat measurements 
and the presence of solitons should be detectable in scattering experiments. 
Another task is the connection, if any, of the soliton phase to the puzzling intermediate phase found in the two-dimensional AFM Kitaev model under the field.
Finally, it would be interesting to study the dynamics of the solitons in a non-equilibrium setting.

\begin{acknowledgments}
This research was supported by NSERC and CIFAR. Computations were performed in
part on the  GPC  and  Niagara  supercomputers  at  the  SciNet  HPC
Consortium. SciNet is funded by: the Canada Foundation for Innovation under the
auspices of Compute Canada; the Government of Ontario;  Ontario  Research  Fund
- Research  Excellence; and the University of Toronto.  Computations were also
performed in part by support provided by SHARCNET (www.sharcnet.ca) and
Compute/Calcul Canada (www.computecanada.ca). Part of the numerical
calculations were performed using the ITensor library~\cite{itensor}.
\end{acknowledgments}

\bibliography{references}
\end{document}